\documentclass[twocolumn,aps,superscriptaddress,pre]{revtex4}

\usepackage{amsmath,amssymb,graphicx}
\usepackage{algorithmic}
\usepackage{enumerate}
\usepackage{times}
\usepackage{color}
\usepackage{soul}
\definecolor{yblue}{rgb}{0.06, 0.3, 0.57}
\usepackage[pdftex]{hyperref}
\hypersetup{colorlinks=true,linkcolor=blue,citecolor=blue,urlcolor=blue}

\begin{document}

\title{Temperature chaos may emerge many thermodynamic states in spin glasses}

\author{Wenlong Wang}
\email{wenlongcmp@scu.edu.cn}
\affiliation{College of Physics, Sichuan University, Chengdu 610065, China}

\begin{abstract}
We present a large-scale simulation of the three-dimensional and mean-field spin glasses down to a very low but finite temperature. We extrapolate pertinent observables, e.g., the disorder-averaged central weight to zero temperature, finding that many thermodynamic states at a finite temperature and two ground states at zero temperature are fully compatible. While the disorder-averaged central weight monotonically decreases with decreasing temperature, this is far from true for individual samples. This motivates us to link this behaviour with the well-known temperature chaos. At an observing temperature, a sample may or may not have pure state coexistence depending on whether it is undergoing temperature chaos, which is a random process. Therefore, temperature chaos is likely responsible for the emergence of many pure states, providing a natural and intuitive explanation for the coexistence of expensive domain-wall excitations and many pure states at the disorder-averaged level. %Finally, we present a crossover picture of the short-range spin glasses.
%exploring the temperature dependence of the spin overlap distribution to very low temperatures for both the EA3D and the SK models. The goal is to argue that a single pair of ground states does not imply a single pair of pure states at a finite temperature.
\end{abstract}

%\pacs{75.50.Lk, 75.40.Mg, 05.50.+q, 64.60.-i}
\maketitle

%\section{Introduction}
\textit{Introduction--} 
Spin glasses are fascinating disordered and frustrated magnets with intriguing properties and diverse applications in biology, computer science, and optimization problems \cite{Young:RMP,book}. Spin glass has a rugged (free) energy landscape, and it is frequently considered as a prototypical complex system. The mean-field Sherrington-Kirkpatrick (SK) model \cite{SK} has an unusual low-temperature phase of many pure states described by the replica symmetry breaking (RSB) \cite{parisi:79,parisi:80,parisi:83}. Here, a pure state refers to a self-sustained thermodynamic state characterized by a time-averaged spin orientational pattern. 

Despite several decades of efforts, it remains an outstanding problem whether the more realistic Edwards-Anderson (EA) spin glass \cite{EA} in three dimensions is qualitatively described by the mean-field theory.
%Despite much research aiming at discriminating which picture is suitable in three (and also four) dimensions, the problem has not been definitely settled. 
The RSB picture \cite{parisi:08,mezard:87} assumes that the mean-field description remains qualitatively correct. On the other hand, the droplet picture \cite{mcmillan:84b,bray:86,fisher:86,fisher:87,fisher:88} based on the domain-wall renormalization group method predicts only a single pair of pure states much like a ferromagnet. The two pictures also differ on the geometrical aspect of excitations (fractals or space-filling) \cite{Wang:Fractal}, and the existence of a spin-glass phase in a weak external magnetic field \cite{almeida:78,Mike:AT}. There are other pictures as well \cite{book}. The applicability of RSB is of broad interest and is related to, e.g., the Gardner transition in structural glasses \cite{charbonneau:14,Mike:GT}. %Here, we focus on the number of pure states, which is a fundamental problem, as a solid answer on one individual property can put stringent constraint on possible theories.

To gain more insights on spin glasses, we have been focusing on the number of pure states, which is a fundamental problem, in recent years. The idea is that a solid answer on one individual property can put stringent constraint on possible theories. Here, we briefly summarize the key progresses in this research direction, including the relevant works of others. 
%There is mounting evidence that the disorder-averaged overlap function is nontrivial (corresponding to many pure states) for the sizes available, which have been steadily growing over time.  
There is mounting evidence that there are many pure states for the sizes available, which have been steadily growing over time. 
To our knowledge, all numerical simulations find a nontrivial central weight [see Eq.~(\ref{I02})] under periodic boundary conditions (PBC) at a typical low temperature, see, e.g., \cite{Enzo:Review,palassini:00,Yucesoy:Delta,Wang:PA,Wang:Delta}.
This is a \textit{numerical fact}. Therefore, to argue against many states, it is necessary that one or several of the conditions have to be altered, e.g., by introducing new statistics or boundary conditions. 
The free boundary condition seemed to support the two-state picture as the central weight decays for small sizes following the $1/L^\theta$ scaling, where $\theta$ is the interface free energy exponent \cite{Katzgraber:FBC}. However, we find that this is a finite-size effect from the surfaces \cite{Wang:FBC} and the central weight of the free and periodic boundary conditions run together for larger sizes, supporting many pure states. 
%This also suggests that the many pure states are genuinely due to droplet excitations rather than topologically protected domain walls trapped within the system.
Second, various statistics were proposed to support the two-state picture, but most of them are not very successful; see, e.g., \cite{Yucesoy:Delta2} and the references therein for a collection of examples. This line of research tells us that it is of paramount importance to conduct contrast simulations for the EA and SK spin glasses if relevant to obtain reliable conclusions. 
%clear interpretations. 
The fraction of centrally peaked instances was a controversial but stimulating statistic \cite{Yucesoy:Delta,billoire:13,Yucesoy:Reply}, but this was also recently clarified \cite{Wang:Delta}. This statistic has a complicated dependence on the system size and the temperature, and it is also sensitive to the fluctuation of the central weight. In addition, suitably modified statistics suggest that the many-state picture is again most likely and coherent \cite{Wang:Delta}.
A work on thermal boundary conditions along with a sample stiffness extrapolation argued against many states \cite{Wang:TBC}. This was partially addressed \cite{Wang:EA4D}, and we shall discuss this again later in this work. 
%To fully resolve this problem a contrast experiment of the SK model should be conducted, which shall be discussed elsewhere. 
Similarly, a number of pioneering works focused on zero temperature, finding a single pair of ground states \cite{Young:GS,Hatano:GS,Itoi:GS}. It is frequently believed that %many pure states at a finite temperature imply many ground states at zero temperature, and 
two ground states at zero temperature implies two pure states at a finite temperature. Recently, we conjectured that a single pair of ground states is a strong support for neither a two-state nor many-state picture \cite{Wang:Delta}. The two ground states is likely due to the relative weights of the low-lying pure states, as even an $O(1)$ excitation is forbidden at $T=0$. Nevertheless, this idea has not been systematically investigated.

The meaning of the number of pure states
%which is related to the presence or absence of RSB, 
requires more discussion at zero temperature to avoid an ambiguous description. One scenario is that there is just a single pair of pure states in the spin-glass phase. Note that this interpretation is in direct conflict with the above finite temperature results. However, the many-state picture is also possible, the key is that the weights of pure states are important here.
%this is also important for a suitable interpretation of the RSB.
%To this end, the meaning of the RSB and the nature of the spin overlap function require further discussion. 
If there are many pure states in the spin-glass phase, it seems likely that the number of pure states grows as the temperature decreases, this is a counting definition. Interestingly, this does not imply that the overlap function $P(q)$ becomes more complicated, because it depends on the weights. Indeed, the central weight actually decreases approximately linearly as more pure states emerge with decreasing temperature \cite{alvarez:10a}.
Therefore, it is helpful to introduce the effective number of pure states $\mathcal{N}_{\mathrm{eff}}=\exp(-\sum_i w_i \log(w_i))$, where $w_i$ is the weight of pure state $i$ at a temperature. In this way, if a pair of them dominates the weight, the system nevertheless manifests as a two-state system as long as equilibrium properties are concerned. The arise of two effective pure states is quite reasonable, considering that $T=0$ is a rather extreme condition. In this setting, it is possible to have two effective pure states, but there are many hidden or inactive pure states with virtually vanishing weights. This picture is appealing because the (free) energy landscape remains highly rugged and spin-glass optimization remains rather challenging, and it is in line with the finite temperature results. %The Ising ferromagnet genuinely has two ground states and also two thermodynamic states in three dimensions. 
Two genuine pure states by nature and two effective pure states by weight are fundamentally different. The former has a trivial $P(q)$ throughout the low-temperature phase, but the latter has a nontrivial $P(q)$ at finite temperatures and it only becomes trivial in the zero temperature limit.
%Therefore, many pure states may well present at $T=0$, but as the weights are highly biased, we may nevertheless find a trivial overlap function. 
Interestingly, the overlap function is not able to fully characterize the nature of the ergodicity breaking, because it also counts the weights.

The main purpose of this work is to show that a single pair of ground states at zero temperature is fully compatible with many pure states at a finite temperature. Here, we simulate both the three-dimensional and mean-field spin glasses down to a \textit{very low} temperature $T=0.01$, which is close to $T=0$. The SK model is simulated for comparison, which is very important as mentioned earlier. The central weight is studied and it is extrapolated to $T=0$, finding that a finite central weight at a finite temperature and a zero central weight at zero temperature are perfectly consistent. The entropy and the link overlap show similar behaviours.

Our work also offers profound insights into another fundamental controversy about the simultaneous presence of $\theta>0$ and a finite central weight, which holds for both the EA and SK spin glasses.
While the disorder-averaged central weight monotonically decreases with decreasing temperature, this is far from true for individual samples. We link this with temperature chaos \cite{mckay:82,bray:87,Hukushima:Chaos,Katzgraber:Chaos,fernandez:13,Wang:TC,Parisi:TC,Berker:clockglass,Zhai:TC} and propose the chaos picture. A particular sample may undergo temperature chaos, and this leads to pure state coexistence in certain temperature intervals. If we take the disorder average at a working temperature, we find a finite central weight. On the other hand, a domain wall can be costly or not for a sample at a working temperature, again depending on whether it is undergoing temperature chaos. Similarly, we would get a finite stiffness exponent if we take the disorder average. We shall discuss the chaos picture in detail in the results section. %Sec.~\ref{results}.

\textit{Computational setup--} 
%\section{Computational setup}
\label{setup}
We study the three-dimensional and the mean-field Ising spin glasses \cite{EA,SK}. The EA spin glass is defined by the Hamiltonian $H = - \sum_{\langle ij \rangle} J_{ij} S_i S_j$, where $S_i=\pm 1$ are Ising spins and the sum is over nearest neighbours on a simple cubic lattice under PBC. For a linear size $L$, there are $N=L^3$ spins. The random couplings $J_{ij}$ are chosen independently from the standard Gaussian distribution with mean zero and variance one. A set of quenched couplings $\mathcal{J}=\{ J_{ij} \}$ defines a disorder sample. The model has a spin-glass phase transition at $T_C \approx 1$ \cite{katzgraber:06}. The mean-field SK spin glass is defined similarly but on a fully connected graph. The Hamiltonian reads $H = - 1/\sqrt{N} \sum_{i<j} J_{ij} S_i S_j$. The transition temperature is exactly known $T_C=1$.
%standard Gaussian distribution $n(0,1)$ with mean zero and variance one.

Our simulation is carried out using the massively parallel population annealing Monte Carlo method \cite{Hukushima:PA,Wang:PA,Amey:PA,Amin:PA,Weigel:UPA}. Population annealing gradually cools a population of $R$ random replicas starting from $\beta=0$ with alternating resampling and Metropolis sweeps until reaching the lowest temperature of interest following an annealing schedule. In a resampling step when the inverse temperature is increased from $\beta$ to $\beta'$, a replica $i$ is copied according to its energy $E_i$ with the expectation number $n_i=\exp[-(\beta'-\beta) E_i]/Q$. Here, $Q=(1/R)\sum_i \exp[-(\beta'-\beta) E_i]$ is a normalization factor to keep the population size approximately the same as $R$. In our simulation, the number of copies is randomly chosen as either the floor or the ceiling of $n_i$ with the proper probability to minimize fluctuations. After the resampling step, Monte Carlo sweeps are applied to each replica at the new inverse temperature $\beta'$. The preliminary simulation parameters are summarized in Table~\ref{table}. Here, we adopt a linear in $\beta$ schedule from $\beta=0$ to $\beta=1$, followed by a linear in $T$ schedule to $T=0.1$ and another linear in $T$ schedule to the lowest temperature $T=0.01$. We apply $100$ temperature steps in the final stage, and the other temperature steps are shared evenly between the first and the second stages.

\begin{table}
\caption{
Preliminary parameters of the population annealing simulation of the EA and SK spin glasses. Here, $N$ is the number of spins, $R$ is the population size, $T_0$ is the lowest temperature simulated, $N_T$ is the number of temperatures used in the annealing schedule, $N_S$ is the number of sweeps applied to each replica after resampling, and $M$ is the number of samples studied. Note that unequilibrated samples were rerun with (much) larger simulation parameters; see the text for details.
\label{table}
}
\begin{tabular*}{\columnwidth}{@{\extracolsep{\fill}} l c c c c c r}
\hline
\hline
Model &$N$ &$R$ &$T_0$ &$N_T$ &$N_S$ &$M$ \\
\hline
EA &$4^3$  &$3.2\times10^5$ &$0.01$  &$201$ &$10$  &$5000$ \\
EA &$6^3$  &$3.2\times10^5$ &$0.01$  &$201$ &$10$  &$5000$ \\
EA &$8^3$  &$1.6\times10^6$ &$0.01$  &$301$ &$10$ &$5000$ \\
EA &$10^3$ &$1.6\times10^6$ &$0.01$  &$301$ &$20$  &$3000$ \\
SK &$4^3$  &$3\times10^5$ &$0.01$  &$201$ &$10$  &$3000$ \\
SK &$6^3$  &$3\times10^5$ &$0.01$  &$201$ &$10$  &$3000$ \\
SK &$8^3$  &$5\times10^5$ &$0.01$  &$301$ &$10$ &$2000$ \\
SK &$10^3$ &$8\times10^5$ &$0.01$  &$301$ &$20$  &$1000$ \\
\hline
\hline
\end{tabular*}
\end{table} 

Population annealing provides intrinsic equilibration measures from the resampling process, such that we can readily examine the thermal equilibration at the level of individual samples. Here, a family name or an index is assigned to each replica in the initial population, and the family names are copied as the replicas are copied. Our equilibration criterion is based on the replica family entropy $S_f = -\sum_i f_i \log (f_i)$, where $f_i$ is the fraction of replicas descended from replica $i$ of the initial population. Intuitively, the family entropy characterizes the diversity of the survival families in the population. %The thermal equilibration is good if the family entropy is sufficiently large, i.e., the survival families in the population are sufficiently diverse. 
We require $S_f \geq \log(100)$ at the lowest temperature for each individual instance \cite{Wang:TBC,Wang:PA}. The unequilibrated instances were rerun with larger parameters until equilibration is reached. It should be noted that the hard instances may require substantially more work than a typical instance. For example, our hardest EA samples of $L=10$ may require $R=2.56\times10^7$ and $N_S=200$ sweeps; cf. the preliminary parameters. 
We simulated a reasonably good number of SK samples, because these long-range samples are expensive to simulate and also they are studied here for the purpose of comparison. 
%To better compare the two models, we simulate the two models with the same number of spins and also identical annealing schedules. 
Finally, our data readily pass the disorder-average based equilibration test of \cite{katzgraber:01} down to the lowest temperature for both models. 
%We check the KY test, they are $[\mean{q_l}]=1+\frac{T[\mean{u}]}{d}$ and $[\mean{q_l}]=1+2T[\mean{u}]$ for the EA and the SK models, respectively.

Our primary observables are the spin overlap distribution $P_{\mathcal{J}}(q)$, the link overlap distribution $P_{\mathcal{J}}(q_l)$, and the entropy $S$. The spin overlap $q$ is defined as:
\begin{equation}
q=\dfrac{1}{N}\sum\limits_i S_{i}^{(1)} S_{i}^{(2)},
\label{q}
\end{equation}
where spin configurations ``(1)'' and ``(2)'' are chosen independently
from the Boltzmann distribution. The central weight of an individual overlap function is defined as:
\begin{align}
    I_{\mathcal{J}}(q_0) = \int_{-q_0}^{q_0} P_{\mathcal{J}}(q) dq.
    \label{I02}
\end{align}
%Here, $q_0$ characterizes the half length of a chosen interval around $q=0$. 
Here, we take $q_0=0.2$ unless otherwise specified. When the subscript is omitted, we refer to the disorder-averaged quantities.
The central weight is a measure of the strength of pure state coexistence.
The link overlap is defined similarly as:
\begin{equation}
q_l=\dfrac{1}{N_B}\sum\limits_{\langle ij \rangle} S_{i}^{(1)}S_{j}^{(1)}S_{i}^{(2)} S_{j}^{(2)},
\label{ql}
\end{equation}
where $N_B$ is the number of bonds. Here, $N_B=3N$ and $N(N-1)/2$ for the EA and SK models, respectively. Importantly, these distribution functions are measured at all temperatures. Finally, the energy $E$, free energy $F$, and the replica family entropy $S_f$ are collected at all temperatures. The free energy can be estimated using the free energy perturbation method \cite{Wang:TBC,Wang:PA}. The entropy $S$ then can be computed from the energy and the free energy as $F=E-TS$.

We expect that the entropy $S$, the link overlap $q_l$, and the central weight $I$ should converge to $\log(2), 1, 0$, respectively, in the $T=0$ limit if there is a single pair of ground states. %They are clearly not so at finite temperatures. 
Here, we systematically study these observables as a function of the temperature and their extrapolation to $T=0$.
The extrapolation is carried out by the cubic polynomial extrapolation. 
%The cubic function is quite flexible, therefore it fits the data well if the interval is reasonably small. 
The errorbar is calculated by the bootstrapping method. We can tune the upper temperature $T_f$ for the fit, i.e., we fit the data for $T<T_f$. It should not be too large, such that the cubic fit works well. On the other hand, it should not be too small for better accuracy. Here, we work with a reasonable value $T_f=0.2$. We have checked that our conclusion is not sensitive to this particular choice. The technique was applied to the three-dimensional Ising ferromagnets with correlated disorder, and it is a reliable method \cite{Wang:Ising3D}.

\begin{figure*}[thb]
\begin{center}
\includegraphics[width=\columnwidth]{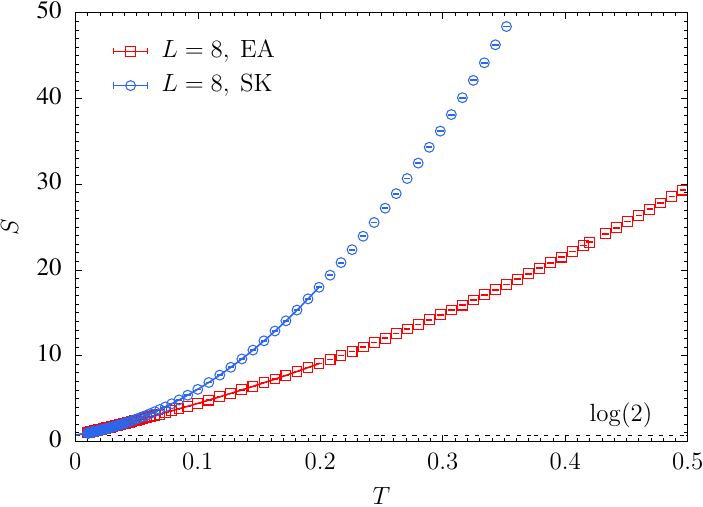}
\put (-22,158) {$(a)$} 
\includegraphics[width=\columnwidth]{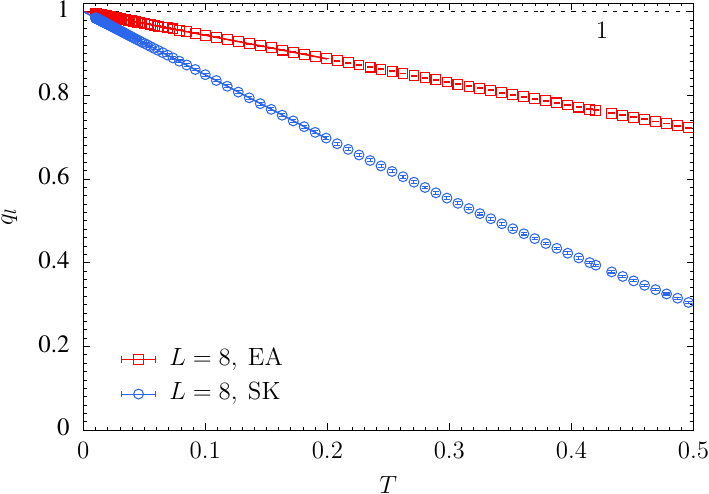}
\put (-21,156) {$(b)$} \\
\includegraphics[width=\columnwidth]{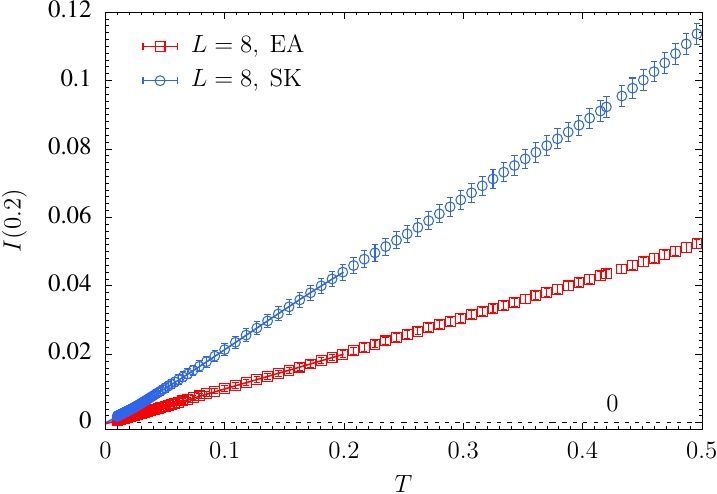}
\put (-22,34) {$(c)$} 
\includegraphics[width=\columnwidth]{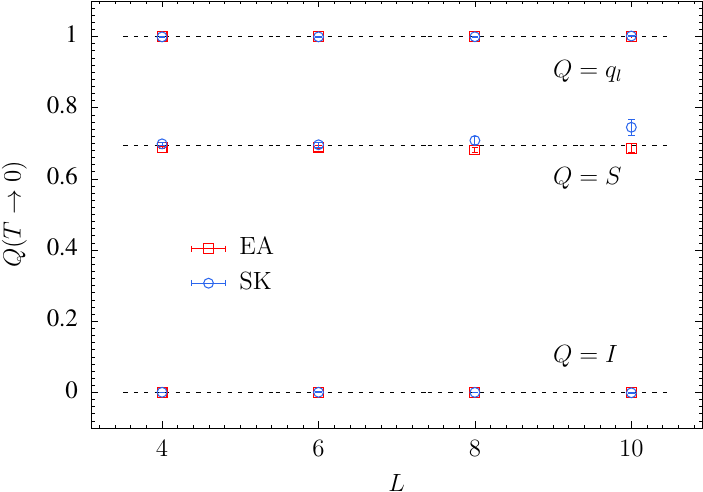}
\put (-17,38) {$(d)$}
\caption{
Temperature evolution of entropy (a), link overlap (b), and central weight (c) for a typical size $L=8$ of both the EA and SK models, along with a cubic fit for $T<0.2$ for extrapolation to $T=0$.
%[panel $(a)$]. 
The extrapolated zero temperature limits (d) are summarized for the various sizes, showing that the three quantities gradually converge to $\log(2), 1$, and $0$ respectively. This crossover behaviour strongly suggests that many pure states at a finite temperature and two ground states at zero temperature are fully consistent.
}
\label{Statistics}
\end{center}
\end{figure*}

\textit{Results--} 
%\section{Results}
\label{results}
The temperature evolution of the disorder-averaged entropy, link overlap, and the central weight along with their zero temperature extrapolations are summarized in Fig.~\ref{Statistics}. First, the detailed results for a typical size $L=8$ are illustrated. Here, a cubic fit is also shown for each data set. It is remarkable that all of these quantities converge to their suitable zero temperature limits of $\log(2), 1, 0$, respectively. For example, the central weight is well finite at finite temperatures, yet it gradually converges to $0$ in the zero temperature limit in an approximately linear manner. %As mentioned earlier, the errorbar of each zero temperature estimation can be obtained by the bootstrapping method. 
The zero temperature estimates of all sizes are summarized in the final panel. The overall data strongly suggest that many pure states at a finite temperature and two ground states at zero temperature are fully compatible, which is a major result of this work.

\begin{figure}[thb]
\begin{center}
\includegraphics[width=\columnwidth]{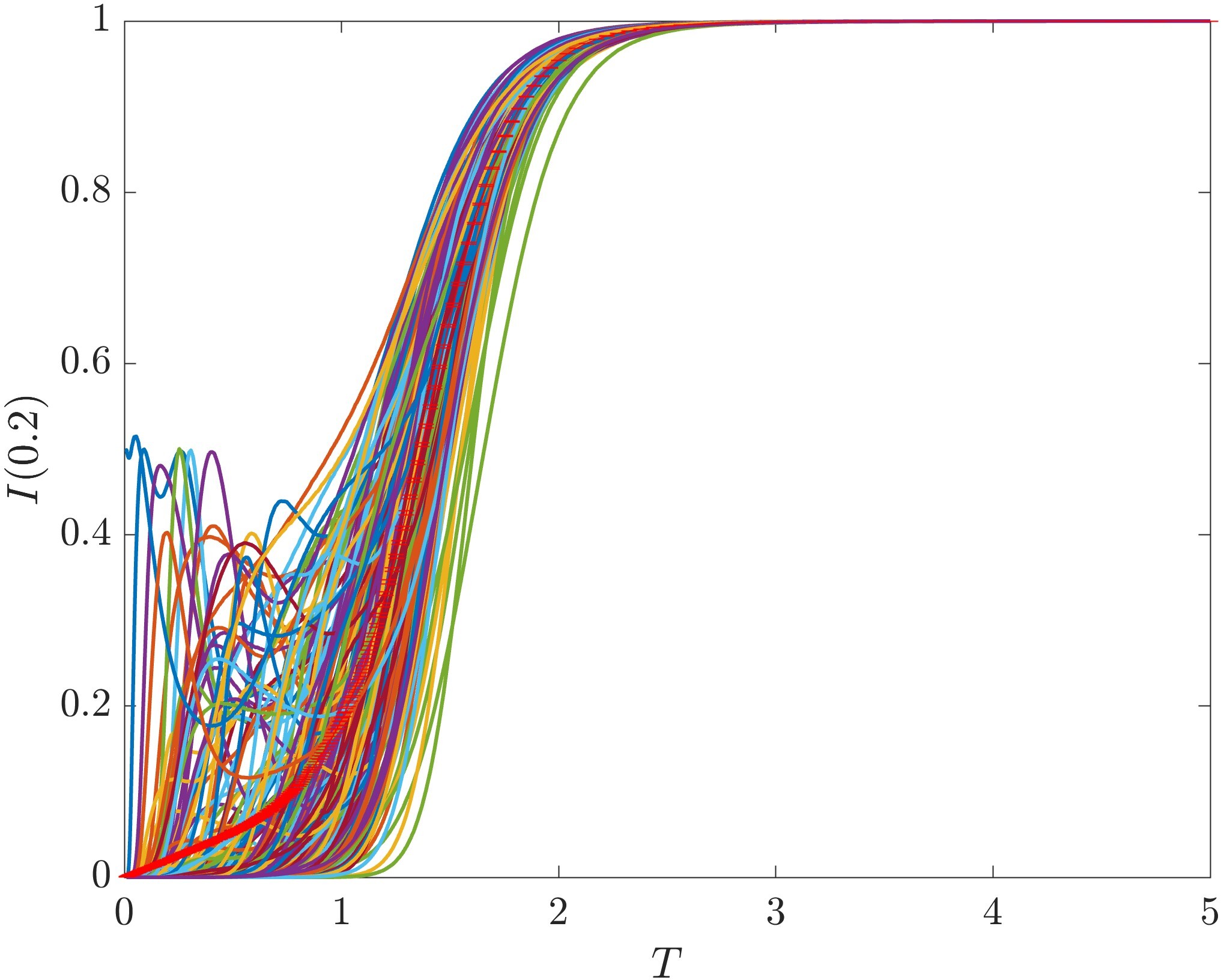}
\caption{
While the disorder-averaged central weight is fully monotonic, this is far from so for many individual samples. Here, only $200$ random samples (EA, $L=8$) are shown for clarity, and the disorder-averaged data are overlaid on top with errorbars. This oscillatory behaviour is also found for the other sizes and the SK model.
}
\label{ITEAL8}
\end{center}
\end{figure}

Before we discuss the number of thermodynamic states, we would like to summarize the following numerical facts:
\begin{enumerate}
    \item The central weight with the PBC at a typical low temperature is nontrivial. 
    \item Most overlap distribution functions are simple while some are nontrivial at a typical low temperature for the sizes available.
    \item The disorder-averaged domain-wall free energy is costly in 3D, i.e., $\overline{\Delta F} \sim L^\theta$, with $\theta>0$, where $\Delta F$ is the free energy difference between the PBC and the anti-periodic boundary condition along the $x$ direction.
    \item Spin glass shows temperature chaos, and the droplet picture describes them well.
\end{enumerate}
These features are also essentially relevant to the SK spin glass, to our knowledge. Particularly, the stiffness exponent $\theta>0$ for the SK model \cite{Wang:KAS,Wang:SKD}. In the following, we link these facts with our chaos picture.
%The disorder-averaged central weight with the PBC at a typical low temperature is approximately a constant function of the system size currently available, which is steadily increasing. 
%particularly the central weight is approximately a constant function of the system size in the spin-glass phase. 
%The simultaneous presence of $\theta>0$ and a finite central weight is not unique to the EA3D model, it also holds for the SK model. The presence of many states in the SK model was not entirely intuitive to us, but it is now intuitive to understand from the perspective of temperature chaos.
%Spin glass chaos; Thermodynamic state chaos; Pure state chaos.
%and bond chaos

The central weight monotonically decreases with decreasing temperature and it converges to $0$ in the zero temperature limit. %While wondering where are the $O(1)$ droplet excitations from, we accidentally realized that 
Intriguingly, there is a quite similar trend in the strength of temperature chaos, which decreases as $T$ decreases, see \cite{Wang:TC,Wang:EA4D} for details. This observation brings us back to the chaos picture, which we proposed earlier, but now we are able to develop it much further \cite{Wang:EA4D}.
Temperature chaos refers to large-scale reorganizations of spin glasses as the temperature is changed.
%Here, we discuss that temperature chaos may emerge many pure states.
For simplicity, we focus on the exchange of weights of the dominant pure states.
If a pure state $\alpha$ dominates at a typical low temperature $T_1$, then another pure state $\gamma$ may become dominate at $T_2 \lesssim T_1$. According to the scaling of temperature chaos, this exchange process takes a temperature interval, which scales as $\delta T\sim 1/L^\zeta$, where $\zeta=d_s/2-\theta$ is the chaos exponent. Here, $d_s$ is the fractal dimension of the droplet or domain wall. In the exchange process, we expect that the weights of the two pertinent pure states are similar or even identical in the relevant temperature interval. Here, we accidentally register a pure state coexistence, which contributes to a nontrivial overlap function. The pure state $\gamma$ remains dominant for a little while before it is ready for the next exchange process, perhaps with the pure state $\delta$, as the temperature decreases. 
It is also typically assumed that the temperature scale between two dominant crossings is also $1/L^\zeta$. Therefore, the system oscillates between a single pure state and pure state coexistence in the process of temperature chaos, and both regions have a sizable share in a temperature interval. In this vein, if we take the disorder average, we naturally find a nontrivial order parameter. This main feature is robust if multiple pure states are exchanging their weights. We mention in passing that there is a scenario to save the droplet picture, if the crossing interval becomes increasingly small relative to the $1/L^\zeta$ interval by an additional factor of $1/L^\theta$. Fortunately, we have already examined this problem earlier \cite{Wang:EA4D}. The crossing interval is in fact $O(1/L^\zeta)$, i.e., the raise and fall of pure state weights is actually quite smooth, despite that the process becomes increasingly more rapidly as the system size increases \cite{Wang:EA4D}.

%While the disorder-averaged central weight is monotonic, the 
The chaos picture predicts that the central weights of individual samples are likely nonmonotonic as they undergo temperature chaos.
Our numerical results confirm this prediction, it is striking that the central weights of many individual samples exhibit oscillatory behaviour in the low-temperature phase, as illustrated in Fig.~\ref{ITEAL8}.
%This can be viewed as the first motivation. We have to look for more evidence for the picture, the nonmonotonic $I(T)$ of some individual samples is a strong one, as we predicted it first and then it is observed.
%It is striking that while the disorder-averaged central weight is monotonic, the central weight of many individual samples is far from so, as illustrated in Fig.~\ref{ITEAL8}. 
%Particularly, the central weight of many samples oscillate in the low-temperature phase. 
This provides strong support to our chaos picture.
%the credibility of 
%This result is very exciting to us, as we realized immediately after seeing it that this is likely temperature chaos and it offers an intuitive picture to essentially eliminate the controversial on the coexistence of a finite $I$ and a finite $\theta$. 
Figure~\ref{ITEAL8} suggests that a sample may contribute to the central weight if it is undergoing temperature chaos, and the contribution oscillates as the temperature decreases in the spin-glass phase. Therefore, at an observing temperature, most overlap distribution functions are simple, but some are nontrivial because these samples are undergoing temperature chaos with pure state coexistence. In the chaos picture, the central weight is not only caused by temperature chaos, but also it is a helpful indicator of the strength of temperature chaos. 
%Therefore, at an observing temperature, most samples are droplet like, but some samples are RSB like because they are undergoing temperature chaos with pure state coexistence.

\begin{figure}[thb]
\begin{center}
\includegraphics[width=\columnwidth]{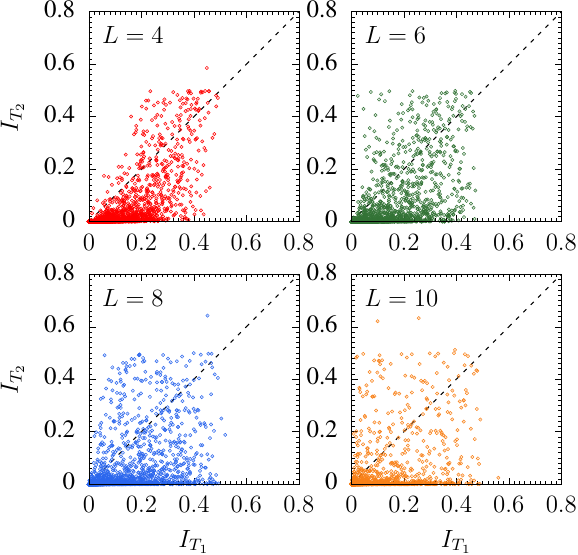}
\put(-24,38) {$(a)$} \\
\includegraphics[width=\columnwidth]{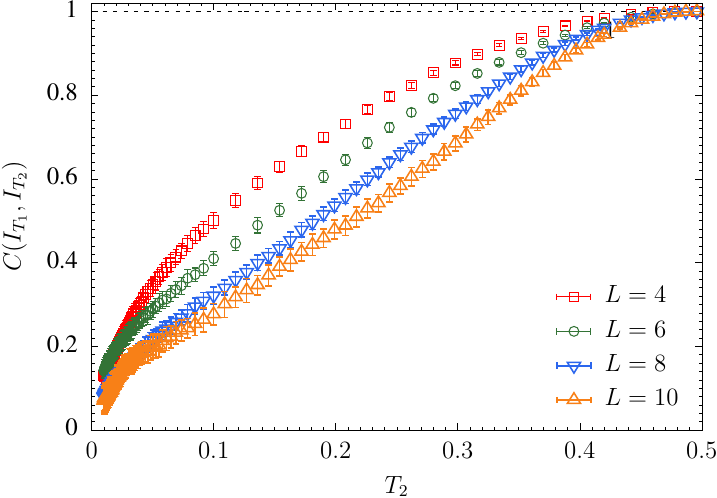}
\put(-209,156) {$(b)$} 
\caption{
\textit{Top panels:} Scatter plot of the central weights of individual samples at $T_1=0.496$ and $T_2=0.208$, showing that the samples contributing to the central weight at two different temperatures are not necessarily the same samples.
\textit{Bottom panel:} The Pearson correlation coefficient shows the decorrelation of the central weights as $T_2$ moves further away from $T_1=0.496$ or as the system size increases. We also find a similar behaviour for the SK model.
}
\label{IT12}
\end{center}
\end{figure}

The chaos picture suggests that the samples that contribute to the central weight at different temperatures are kinetic rather than static, i.e., they are not necessarily the same samples.
%at a temperature $T_1$ and another temperature $T_2$ are not necessarily the same samples. 
In this scenario, we expect that the central weights of two different temperatures $I_{T_1}$ and $I_{T_2}$ should decorrelate as $\Delta T=|T_1-T_2|$ increases or alternatively as $L$ increases. To this end, a scatter plot of them for $T_1=0.496$ and $T_2=0.208$ is shown in Fig.~\ref{IT12}. It is reasonably obvious that they become increasingly decorrelated as $L$ increases. Note that temperature chaos is stronger for larger sizes. The Pearson correlation coefficient is also shown, here $T_1=0.496$ and $T_2$ is tuned for various sizes. It is evident that the correlation decreases as $T_2$ moves further away from $T_1$ or as the system size increases until they overlap together in the tail, where the correlation is weak and the statistical error is larger. We also find a similar behaviour for the SK model, which is omitted here for simplicity.

The chaos picture naturally explains the coexistence of a finite central weight and a finite stiffness exponent $\theta$, which has been a longstanding puzzle in spin glasses. Here, the droplet or domain-wall free energy is $O(L^\theta)$ or $O(1)$ in the temperature interval without or with pure state coexistence, respectively.
%While some domain walls are not expensive due to temperature chaos, most domain walls are nevertheless expensive. 
The tricky part is that the scaling of $\theta$ is dominated by the expensive domain walls. Therefore, at an observing temperature, the small number of chaotic samples yield a finite central weight and most non-chaotic samples yield a finite stiffness exponent. These apparently contradictory results arise due to the two ``types'' of samples, and also because we are taking the \textit{disorder average} of the samples. Of course, there are no genuinely two types of samples, this depends on temperature chaos and again the classification evolves with the temperature. Therefore, there is no real contradiction between these two facts, and this is likely a statistical artifact.
%an illusion of statistics.
%Statistical illusion; Statistical artifact；

The two ``classes'' of samples picture asserts that the scaling of $\Delta F$ only works for the bulk of its distribution, particularly the tail of the distribution, i.e., the scaling may not work when $\Delta F \sim O(1)$. Indeed, we expect that the $O(1)$ excitations are always present at the disorder-averaged level due to temperature chaos. 
%It is known that the estimate of $\theta$ decreases slightly when the temperature departs further from $T=0$ for the same range of sizes. Our picture is in line with this observation, as more $O(1)$ $\Delta F$ values are available at higher temperatures. Note that temperature chaos is stronger at higher temperatures. When they are combined with the $L^\theta$ scaling of the expensive excitations, this tends to underestimate the $\theta$ exponent. 
Fortunately, we have the $\Delta F$ data available from an earlier work, we simulated $5000$ samples for each of $L=4, 6, 8, 10$ down to $T=0.2$ and the ground states are also available at $T=0$ \cite{Wang:Delta}. Interestingly, the cumulative distribution function (CDF) of $|\Delta F|$ of various sizes does show a clear scaling in the tail, but they also overlap quite well in a good interval when the CDF is below about $0.05$. This feature is quite robust against temperature changes. To this end, we estimate $\theta=0.180(16),0.217(15),0.022(53)$ at $T=0.2$ for all the data, the largest $5\%$ and the smallest $5\%$ of the data, respectively. Similarly, the corresponding estimates are $\theta=0.185(16),0.219(15),0.078(55)$ at $T=0$ using the ground state energies. Note that the tail yields a larger exponent than the overall scaling, in line with \cite{Wang:TBC}. Particularly, the fitted exponents from the head of the distribution are virtually $0$.
%Particularly, if we take the largest $10\%$ excitation (free) energies, we obtain a good estimate of $\theta$. However, if we take the smallest $10\%$ of them, the estimate is essentially $0$, in line with the chaos picture. 
Therefore, there is evidence that the $L^\theta$ scaling does not hold for the $O(1)$ excitation free energies. The chaos picture also strengthens the interpretation \cite{Wang:EA4D} to the sample stiffness extrapolation using thermal boundary conditions \cite{Wang:TBC}. If a fraction of samples never becomes stiff, i.e., boundary coexistence (similar to pure state coexistence) remains in the thermodynamic limit, then a finite central weight is obtained as observed. A systematic discussion of the domain-wall free energy distribution and the pertinent scaling should be presented in a different work.

\textit{Summary and outlook--} 
%\section{Summary and outlook}
\label{conclusions}
In this work we carried out a large-scale simulation of the EA and SK spin glasses down to a very low temperature $T=0.01$. We measured several statistics as a function of the temperature, particularly the central weight, and they are extrapolated to the zero temperature limit. A crossover behaviour is found for all of these statistics, showing that many pure states at a finite temperature and two ground states at zero temperature are fully consistent.

We also developed the chaos picture, linking many numerical facts of spin glasses. In this picture, an individual sample exhibits pure state coexistence or not depending on its process of temperature chaos. Similarly, the droplet or domain-wall free energy may be costly or not, again depending on its temperature chaos. Therefore, at an observing temperature, we find a nontrivial overlap function and a finite stiffness exponent at the disorder-averaged level.
%This picture predicts two effective classes of samples, and has many observable consequences. 
Future work should test the chaos picture further. For example, the distribution of $\Delta F$ should be systematically studied. We also predict that the central weight of the thermal boundary conditions is likely larger than that of the PBC even in the thermodynamic limit. A simulation of spin glasses to larger sizes and also in higher dimensions is also particularly interesting. 
Research efforts along some of these directions are currently in progress and will be reported in future publications.

%offer a resolution 
%With one more controversy solved and supporting (again) many pure states and the difficulties of the two state picture mentioned above, it appears that a coherent picture of many pure states is emerging and that the spin glass balance is significantly tiling towards many pure states by recent simulations. Further explorations such as the SK spin glass in thermal boundary conditions \cite{Wang:TBC} are currently in progress and will be published in the future.

%\acknowledgments 
\textit{Acknowledgments--} 
We gratefully acknowledge supports from the National Science Foundation of China under Grant No. 12004268, and the Fundamental Research Funds for the Central Universities, China.
We thank the Emei cluster at Sichuan University for providing HPC resources.
%We thank J. Machta, M. Weigel, and B. Yucesoy for helpful discussions. W.W.~acknowledges support from the Swedish Research Council Grant No.~642-2013-7837, and the Goran Gustafsson Foundation
%for Research in Natural Sciences and Medicine, and the Fundamental Research Funds for the Central Universities, China. M.W.~acknowledges
%support from the Swedish Research Council Grant No.~621-2012-3984.
%The computations were performed on resources provided by the Swedish
%National Infrastructure for Computing (SNIC) at the National
%Supercomputer Centre (NSC), and the High Performance Computing Center
%North (HPC2N), and the Emei cluster at Sichuan university.

\bibliography{Refs}

\end{document}